\documentclass[%
 aps,
 prapplied,%
 amsmath,amssymb,
 preprint,%
superscriptaddress,
]{revtex4-2}

\usepackage{graphicx}% Include figure files
\usepackage{dcolumn}% Align table columns on decimal point
\usepackage{bm}% bold math
\usepackage{here}
\usepackage{comment}
\usepackage{color}
\usepackage{hyperref}
\begin{document}

\preprint{Surface phonons, Morita, ver.2}

\title{Surface phonons limit heat conduction in thin films}% Force line breaks with \\

\author{Michimasa Morita}
\affiliation{Department of Mechanical Engineering, The University of Tokyo, 7-3-1 Hongo, Bunkyo, Tokyo, 113-8656, Japan}
\author{Takuma Shiga}
\email{shiga@photon.t.u-tokyo.ac.jp}
\affiliation{Department of Mechanical Engineering, The University of Tokyo, 7-3-1 Hongo, Bunkyo, Tokyo, 113-8656, Japan}
\affiliation{Japan Science and Technology Agency, PRESTO, 4-1-8 Honcho, Kawaguchi, Saitama, 332-0012, Japan}

\date{\today}% It is always \today, today,
% but any date may be explicitly specified

\begin{abstract}
Understanding microscopic heat conduction in thin films is important for nano/micro heat transfer and thermal management for advanced electronics. As the thickness of thin films is comparable to or shorter than a phonon wavelength, phonon dispersion relations and transport properties are significantly modulated, which should be taken into account for heat conduction in thin films. Although phonon confinement and depletion effects have been considered, it should be emphasized that surface-localized phonons (surface phonons) arise whose influence on heat conduction may not be negligible due to the high surface-to-volume ratio. However, the role of surface phonons in heat conduction has received little attention thus far. In the present work, we performed anharmonic lattice dynamics calculations to investigate the thickness and temperature dependence of in-plane thermal conductivity of silicon thin films with sub-10-nm thickness in terms of surface phonons. Through systematic analysis of the influences of surface phonons, we found that anharmonic coupling between surface and internal phonons localized in thin films significantly suppresses overall in-plane heat conduction in thin films. We also discovered that specific low-frequency surface phonons significantly contribute to surface--internal phonon scattering and heat conduction suppression. Our findings are beneficial for the thermal management of electronics and phononic devices and may lead to surface phonon engineering for thermal conductivity control. 
\end{abstract}

\maketitle
\section{Introduction}
Heat conduction analysis of low-dimensional materials, such as thin films, nanowires, and superlattices, is important for nano/micro heat transfer and thermal management for advanced microelectronics \cite{Cahill_review,Nomura_review}. Heat conduction in thin films has been extensively investigated, as the reduced thermal conductivity of thin films leads to poor heat dissipation of electronics \cite{Pop_review}. The Fuchs--Sondheimer (FS) model \cite{Fuch_1938,Sondheimer_1952} has been applied to phonon transport in thin films and has been demonstrated to be valid for reproducing thermal conductivity experiments for thicknesses above 20 nm \cite{Cuffe_2015,Jain_2016}. However, as the thickness of a thin film is comparable to or shorter than a characteristic phonon wavelength, modulation of the phonon dispersion relation and transport properties arises from the low dimensionality \cite{Heino_2007}. Therefore, phonon transport properties in bulk materials, usually input into the FS model, are not valid for describing heat conduction in sub-10-nm-thick films. 

Phonon transport in sub-10-nm-thick thin films has been investigated in various studies. For instance, Neogi and Donadio \cite{Neogi_Donadio_2015} performed molecular dynamics simulations for free-standing silicon thin films with ($2~\times~1$) surface reconstruction and demonstrated that the in-plane thermal conductivity of thin films and its thickness dependence are significantly different from the FS model. Fu \textit{et al.} \cite{Fu_2020} applied anharmonic lattice dynamics to silicon thin films with thicknesses in the range of 1--5 nm and found that the in-plane thermal conductivity of thin films with thicknesses below 2 nm is insensitive to the thickness. In these calculations, although an empirical potential was used to describe the interatomic interactions between silicon atoms, the findings were not affected by the choice of force field. Wang \textit{et al.} \cite{Wang_2019} performed first-principles-based anharmonic lattice dynamics calculations for silicon thin films with thicknesses of 0.94 nm and 1.48 nm and observed a similar thickness dependence. 

Some studies have also investigated how surface roughness affects heat conduction in thin films. Neogi and Donadio \cite{Neogi_Donadio_2015} and Wang \textit{et al.} \cite{Wang_2019} demonstrated that surface roughness significantly reduces thermal conductivity. In addition, they found that the magnitude of the reduction and thickness dependence are consistent with experiments \cite{Asheghi_1997,Ju_1999,Ju_2005,Liu_2005}. Neogi \textit{et al.} \cite{Neogi_etal_2015} also demonstrated that silicon oxide layers at the surface reduce the thermal conductivity of thin films, which is caused by the reduction of the group velocity by localized vibrational modes inside the silicon oxide layers \cite{Xiong_2017}.

In the sub-10-nm thickness regime, a high surface-to-volume ratio leads to strong coupling of surface structures and heat conduction, which opens a new avenue for heat conduction control using nanoengineered surfaces \cite{Neogi_Donadio_2020}. However, it is still worth investigating the intrinsic mechanism of heat conduction in thin films without surface roughness. For such ultrathin films, phonon depletion and confinement effects \cite{Turnery_2010,Wang_2014} have been considered in the literature. However, these effects are not sufficient to explain the reduced thermal conductivity and thickness dependence of the thermal conductivity of ultrathin films. For a comprehensive understanding, it is necessary to consider surface-localized phonons (i.e., surface phonons), which arise in ultrathin films. Similar to vibrational modes localized in surface oxide layers, surface phonons are likely to suppress heat conduction in thin films. Therefore, we evaluated how surface phonons influence heat conduction in free-standing silicon thin films with sub-10-nm thickness by performing anharmonic lattice dynamics calculations. 

\section{Methods}
\begin{figure}[b]
\centering
\includegraphics[width = 0.8\columnwidth]{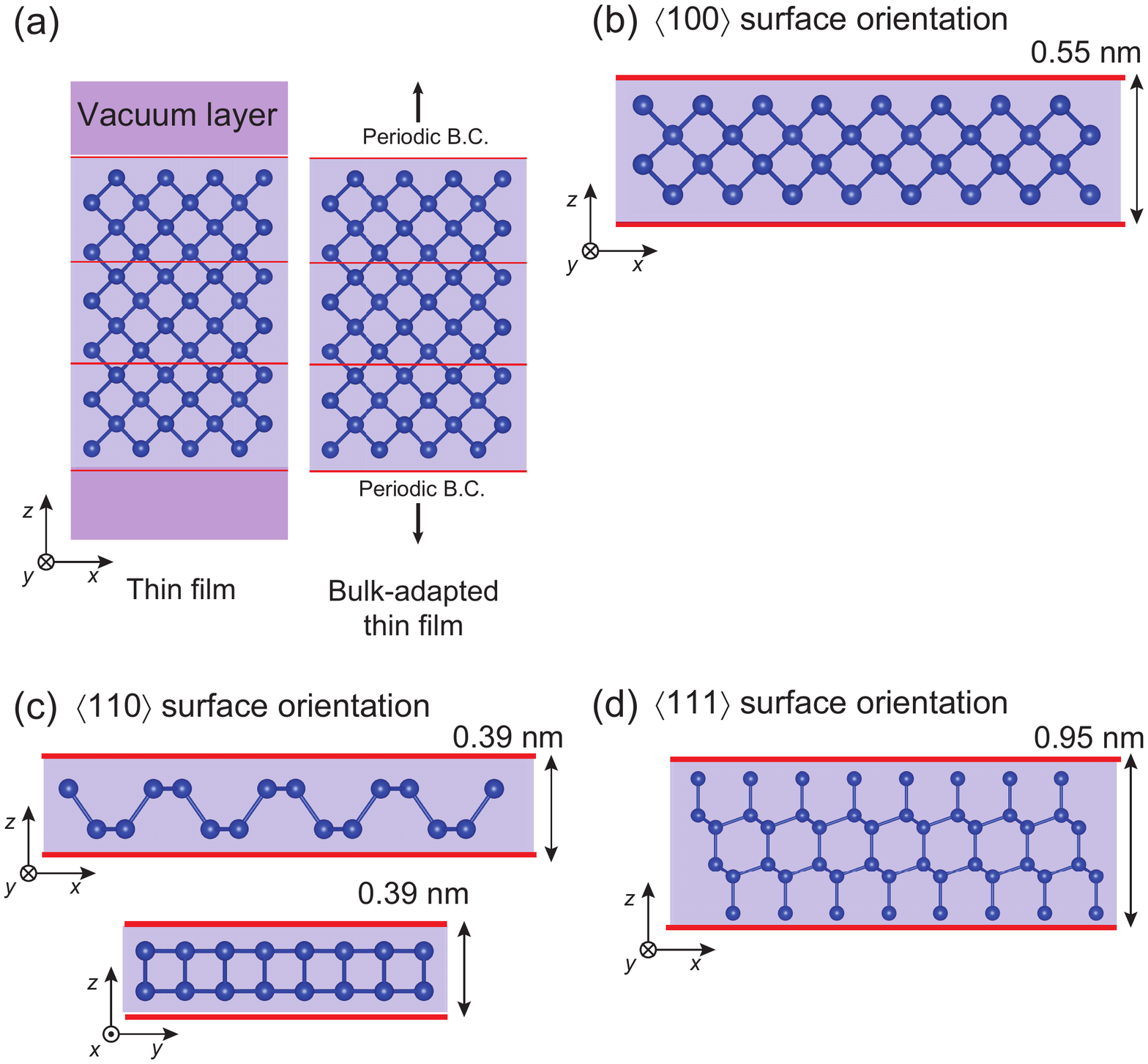} 
\caption{(a) Schematic diagram of thin films. (b)--(d) Unit cell of the film for each surface orientation. Due to the anisotropy of the $\langle 110 \rangle$ surface orientation, two images for the $x$- and $y$-directions are presented.}
\label{fig:1}
\end{figure}

We considered silicon thin films with $\langle 100 \rangle$, $\langle 110 \rangle$, and $\langle 111 \rangle$ surface orientations. A unit cell for each surface orientation is illustrated in Fig. \ref{fig:1}. The lattice constants of the three surface orientations were 0.55 nm ($\langle 100 \rangle$), 0.39 nm ($\langle 110 \rangle$), and 0.95 nm ($\langle 111 \rangle$), respectively (Figs. \ref{fig:1}(b)--(d)). A thin film with a given thickness was modeled by stacking unit cells along the $z$-direction perpendicular to the surface, sandwiched by vacuum layers at the top and bottom surfaces, as illustrated in Fig. \ref{fig:1}(a). Periodic boundary conditions were applied to the $x$- and $y$-directions. 

First-principles calculations can be performed to obtain the interatomic force constants (IFCs) required to calculate phonon transport properties \cite{Broido_2007,Esfarjani_2008,Esfarjani_2011}. However, because the approximately 10-nm-thick film considered in this work included several hundreds of atoms, the use of first-principles calculations was not practical. Therefore, we employed the optimized Stillinger--Weber (SW) potential \cite{SW,Lee_2012,Lee_2012_2}. Lee and Hwang \cite{Lee_2012,Lee_2012_2} adjusted the parameter set of the SW potential using density functional theory calculations to reproduce the phonon dispersion relations and thermal conductivity of bulk silicon obtained from experiments. In our calculations, we chose the parameter set obtained by density functional theory calculations with the generalized gradient approximation. Although surface reconstruction and changes in the bond lengths of surface atoms generally occur \cite{Rosei_2004}, structural relaxation in the thin films with the optimized SW potential did not significantly change the bond lengths of surface atoms (less than 0.1\%) nor produce surface reconstruction. For comparison with the bulk-adapted model \cite{Kress,Allen_1971_1,Allen_1971_2} and to investigate the effect of surface phonons on overall heat conduction in thin films, we studied thin films without surface reconstruction. Furthermore, to enhance the influence of surface phonons on the transport properties, the outermost surface atoms of the $\langle 111 \rangle$ surface orientation, which is usually unstable, were retained. 

Since three--phonon scattering is dominant in thermal resistance, harmonic and third-order anharmonic IFCs were considered, and the interaction ranges were set to the second-nearest neighbors. We calculated the in-plane thermal conductivity of free-standing thin films by solving the phonon Boltzmann transport equation under the single-mode relaxation time approximation. In the calculation of the relaxation times for three--phonon scattering, we accounted for both normal and Umklapp processes \cite{Ziman,Srivastava}. To investigate the intrinsic effects of surface phonons, we did not include the effect of isotope scattering \cite{Tamura}. The Dirac delta function associated with energy conservation in three--phonon scattering was approximated by a Lorentzian with linewidth $\varepsilon$. In the present work, we chose $\varepsilon$ = 10 cm$^{-1}$ and a $20~\times~20$ uniform reciprocal mesh in the two-dimensional first Brillouin zone for calculating the transport properties, which ensured the convergence of thermal conductivity (details are provided in Appendix \ref{appendix:A}). We used the ALAMODE package for all IFCs and anharmonic lattice dynamics calculations \cite{Tadano}.

\section{Results and Discussion}

\begin{figure}[h]
\centering
\includegraphics[width = 0.8\columnwidth]{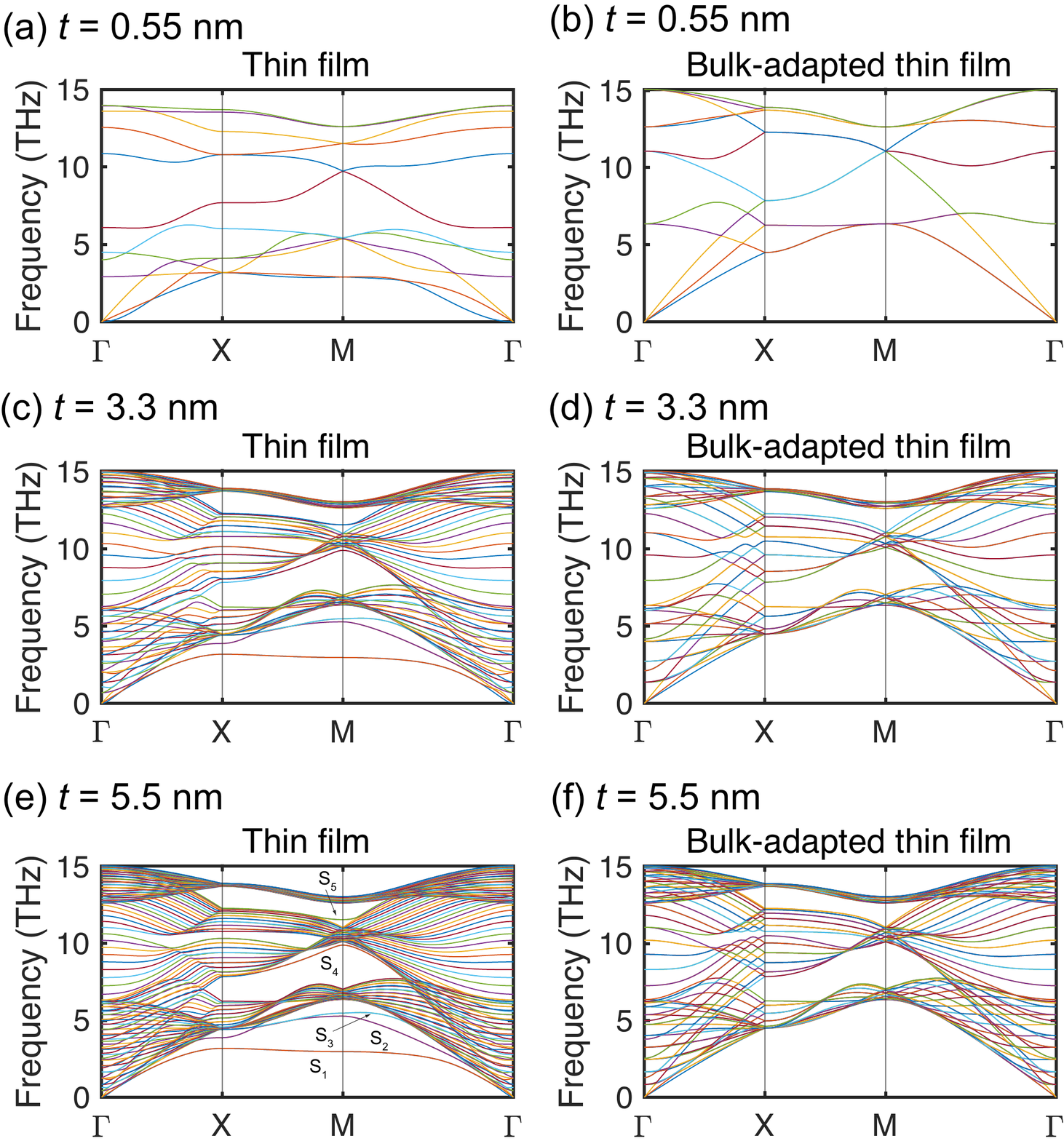}
\caption{(a), (c), (e) Phonon dispersion relations of $\langle 100\rangle$ surface-oriented thin films for three thicknesses ($t$) of 0.55 nm, 3.3 nm, and 5.5 nm, respectively. (b), (d), (f) Phonon dispersion relations calculated from the bulk-adapted method \cite{Kress,Allen_1971_1,Allen_1971_2} for three thicknesses of 0.55 nm, 3.3 nm, and 5.5 nm, respectively.}
\label{fig:2}
\end{figure}

Figures \ref{fig:2}(a), (c), and (e) display the calculated phonon dispersion relations of the $\langle 100\rangle$ surface orientation for three thicknesses. The thinnest film (Fig. \ref{fig:2}(a)) has an out-of-plane acoustic phonon mode whose angular frequency is proportional to the square of the wavevector near the zone center. This feature has been observed in two-dimensional materials \cite{Balandin}. As the thickness increases, the wavevector dependence becomes linear, exhibiting a form of three-dimensional vibrational modes inside the film. Another remarkable feature in the dispersion relations is the presence of isolated phonon modes. These isolated phonon modes can be readily observed, even in relatively thick films, corresponding to surface phonons. In the 5.5-nm thin film (Fig. \ref{fig:2}(e)), there are five surface phonons, labeled S$_1$--S$_5$. Eigenvector analysis reveals that S$_1$ and S$_2$ are in-plane and out-of-plane surface phonon modes, respectively. Surface phonons can also be detected by the bulk-adapted method \cite{Kress,Allen_1971_1,Allen_1971_2}, in which perturbations of harmonic IFCs are eliminated among surface atoms by applying periodic boundary conditions in the direction perpendicular to the surface. A simple comparison between the dispersion relations allows us to find isolated phonons in the low-frequency regime (Figs. \ref{fig:2}(c)--(f)). 

\begin{figure}[h]
\centering
\includegraphics[width = 0.8\columnwidth]{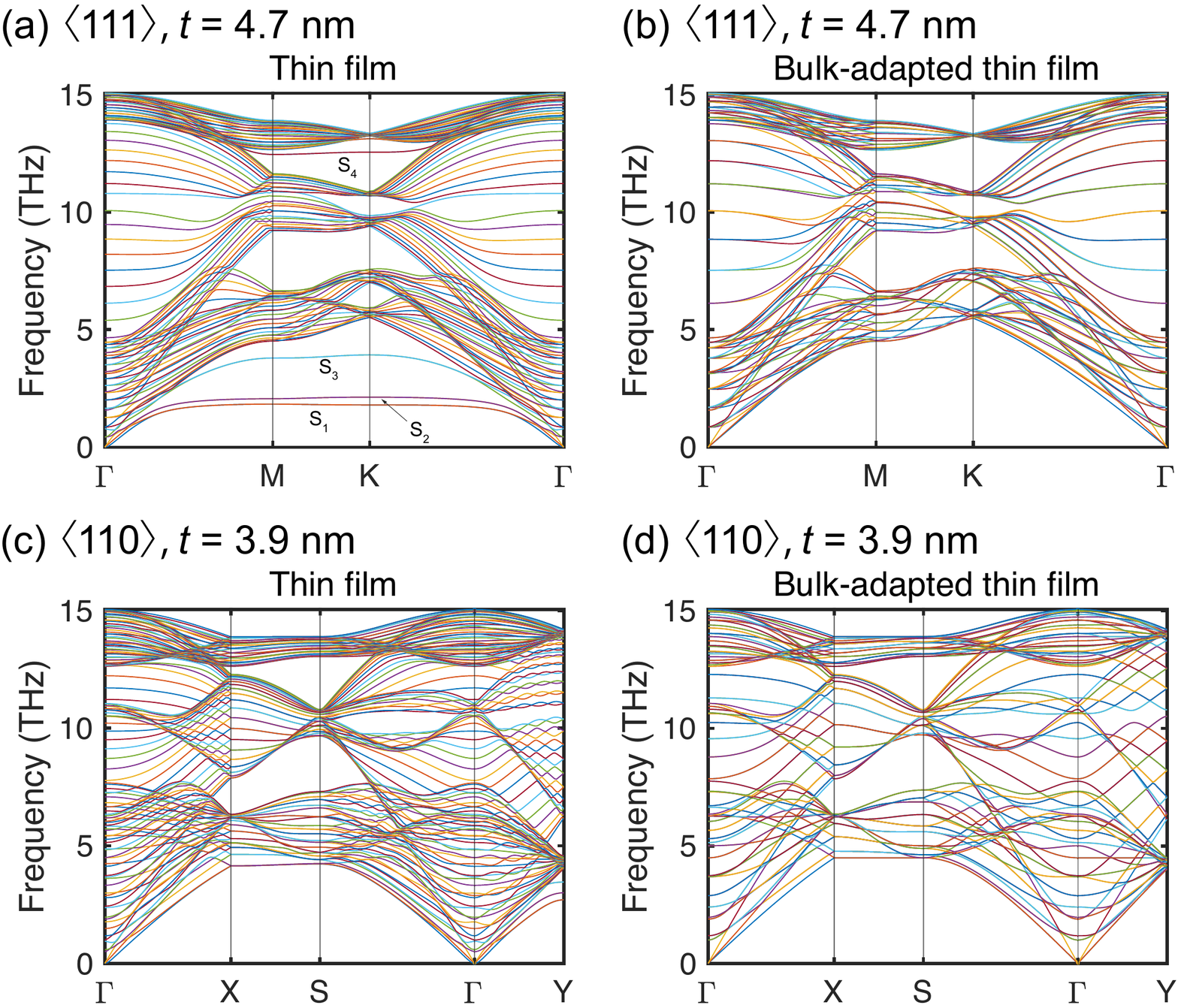}
\caption{Phonon dispersion relations of (a), (c) thin films and (b), (d) bulk-adapted thin films of different thicknesses ($t$) for $\langle 111\rangle$ and $\langle 110\rangle$ surface orientations.}
\label{fig:3}
\end{figure}

For the $\langle 111\rangle$ surface orientation, the $\langle 111\rangle$ surface-oriented thin film also has surface phonon modes (Figs. \ref{fig:3}(a),(b)), although the number and frequencies of the surface phonon modes are different from those of the $\langle 100\rangle$ surface orientation. In contrast, for the $\langle 110\rangle$ surface orientation, surface phonons cannot be identified from the dispersion relations based on comparison with the bulk-adapted method (Fig. \ref{fig:3}(c),(d)), while the frequencies of the low-frequency acoustic modes in the $\Gamma$--Y line are slightly reduced. The characteristics of surface phonons are strongly dependent on the surface orientation, which is due to the coordination number of the outermost surface atoms. Whereas the coordination number of atoms in a thin film is four, that of the outermost surface atoms is one, two, and three for the $\langle 111\rangle$, $\langle 100\rangle$, and $\langle 110\rangle$ surface orientations, respectively. Thus, the harmonic IFC perturbations are the largest (smallest) for the $\langle 111\rangle$ ($\langle 110\rangle$) surface orientation, resulting in discrepancies in the extent of isolation of the surface phonons. 

\begin{figure}[h]
\centering
\includegraphics[width = 0.5\columnwidth]{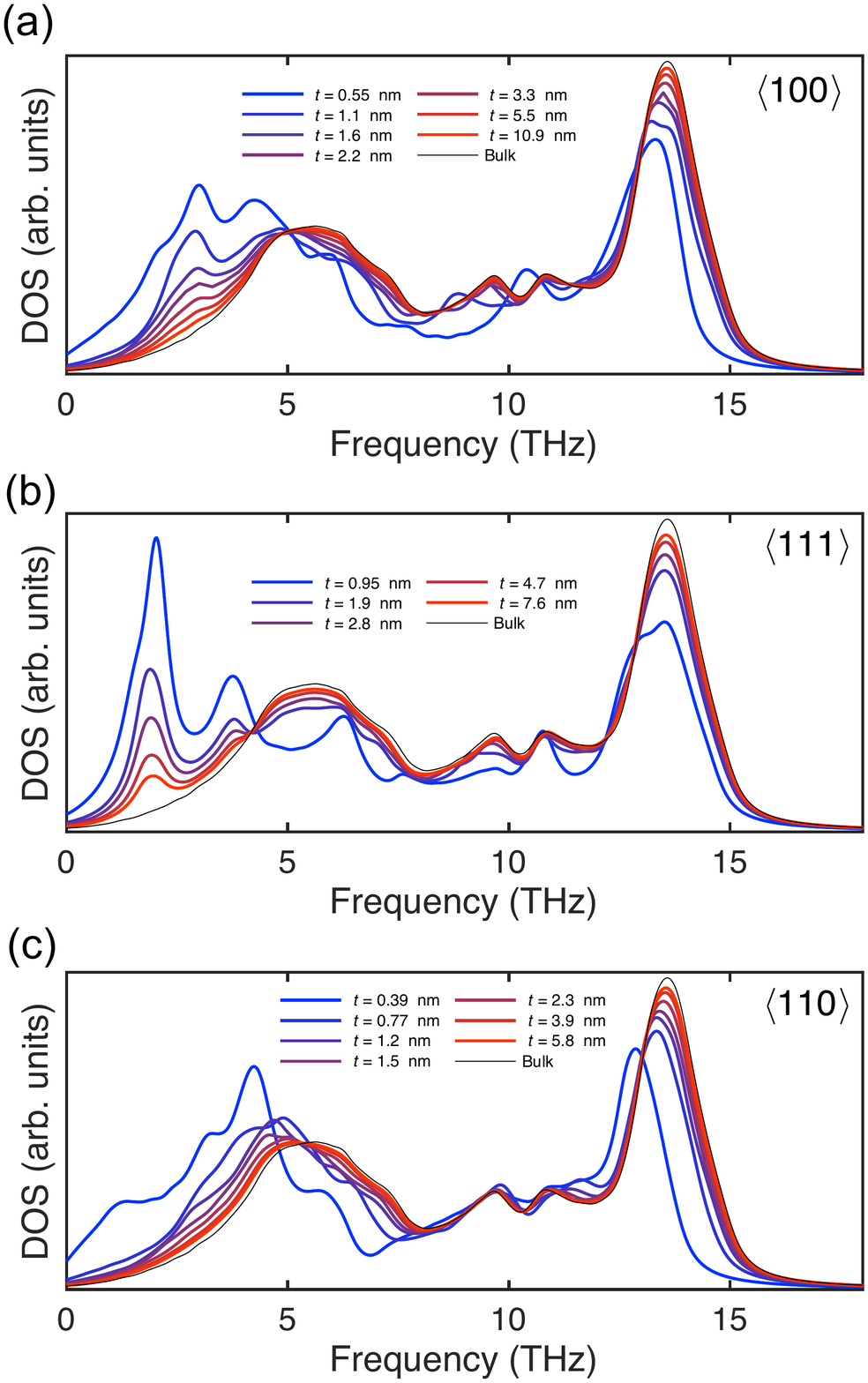} 
\caption{Frequency-dependent phonon density of states (DOS) for (a) $\langle 100\rangle$, (b) $\langle 111\rangle$, and (c) $\langle 110\rangle$ surface-oriented thin films of different thicknesses. The DOS is normalized by the total number of phonons for each surface orientation and thickness.}
\label{fig:4}
\end{figure}

Figure \ref{fig:4} displays the frequency-dependent density of states (DOS) of the three surface orientations. Several characteristic peaks of bulk silicon (represented as solid black lines) are observed for a relatively thick film. Side peaks sensitively changing to the thickness can be seen, particularly for the $\langle 100\rangle$ and $\langle 111\rangle$ surface orientations (Figs. \ref{fig:4}(a),(b)). The frequencies of these side peaks correspond to those of the S$_1$ surface phonon mode and S$_1$--S$_3$ surface phonon modes for the $\langle 100\rangle$ and $\langle 111\rangle$ surface orientations, respectively. The large magnitude of the side peak at approximately 2 THz for the $\langle 111\rangle$ surface orientation is attributed to the flat dispersions of S$_1$ and S$_2$ surface phonon modes whose frequencies are close to each other. Since the number of surface phonon modes is nearly independent of the thickness, the magnitudes of the side peaks characterized by surface phonons decrease monotonically as the thickness increases. For the $\langle 110\rangle$ surface orientation, side peaks cannot be observed due to the absence of distinct isolated phonon modes in the dispersion relation. However, except for the thinnest films, DOS spectra in the frequency of 2--5 THz slightly change based on the thickness, which is due to the modulation of the dispersion relations. 

To quantify the modulation of the phonon dispersion relations and DOS, we calculated the Debye temperature of the thin films. Volumetric heat capacity is given by \cite{Srivastava}
\begin{equation}
	\displaystyle C_{v}=k_\mathrm{B}\sum_{\bm{q}s}\left(\frac{\hbar\omega(\bm{q}s)}{k_\mathrm{B}T}\right)^{2}\frac{\displaystyle \exp\left(\frac{\hbar\omega(\bm{q}s)}{k_\mathrm{B}T}\right)}{\displaystyle \left[\exp\left(\frac{\hbar\omega(\bm{q}s)}{k_\mathrm{B}T}\right)-1\right]^{2}},
    \label{eq:heat_capacity}
\end{equation}
where $k_\mathrm{B}$, $T$, and $\hbar\omega(\bm{q}s)$ are the Boltzmann constant, temperature, and energy of the phonon with wavevector $\bm{q}$ and polarization $s$, respectively. Using the Debye approximation, the volumetric heat capacity can also be expressed as 
\begin{equation}
	\displaystyle C_{v}^\mathrm{D}=9Nk_\mathrm{B}\left(\frac{T}{\Theta_\textrm{D}}\right)^{3}\int_{0}^{\Theta_\mathrm{D}/T}\frac{x^{4}e^{x}}{(e^{x}-1)^{2}}dx,
	\label{eq:heat_capacity_Debye}	
\end{equation}
where $N$ denotes the number of atoms in a primitive unit cell. We obtained the Debye temperature ($\Theta_\mathrm{D}$) to match $C_{v}^\mathrm{D}$ with $C_v$ using the Newton--Raphson method. 

\begin{figure}[h]
\centering
\includegraphics[width = 0.8\columnwidth]{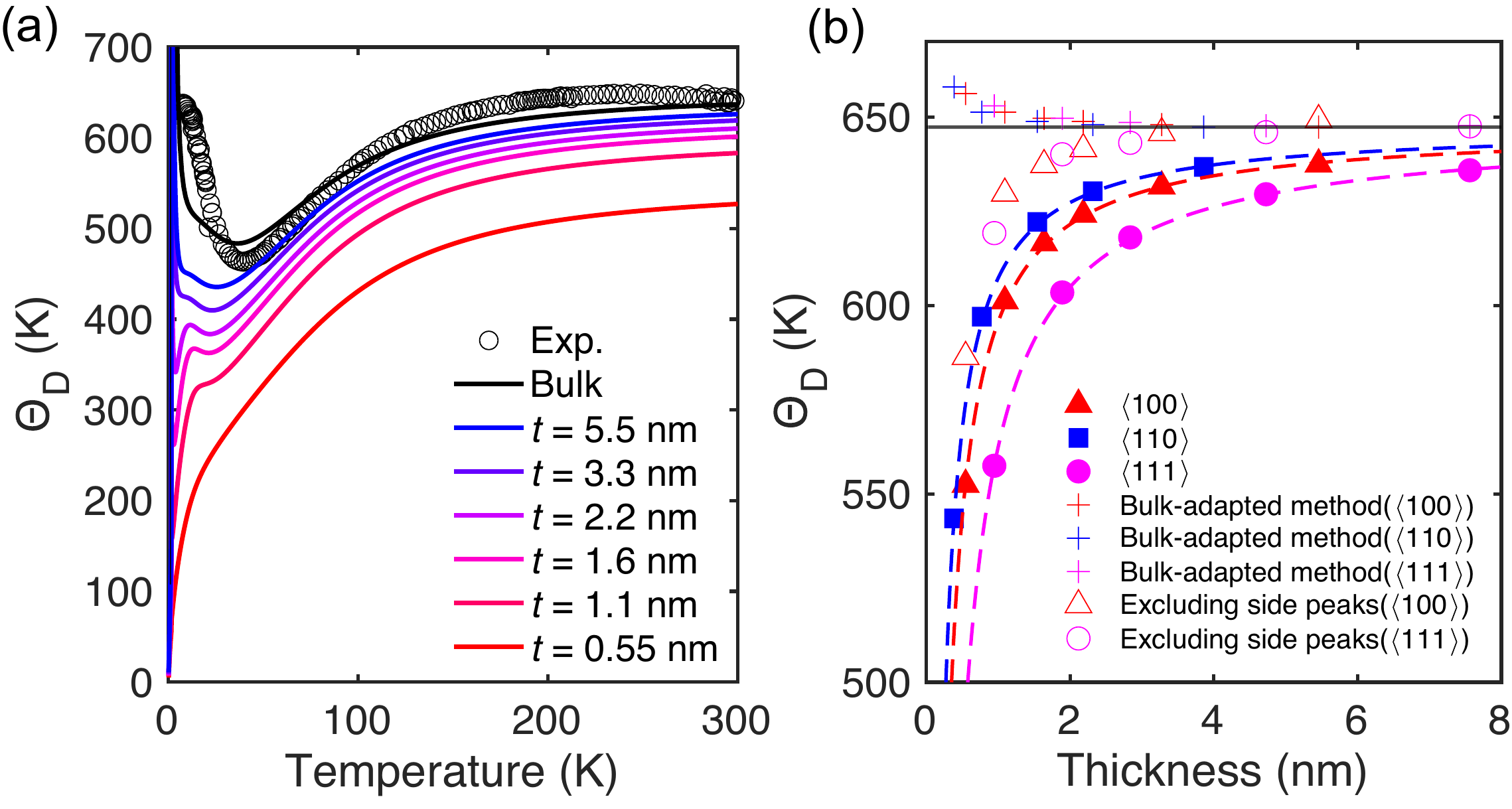} 
\caption{(a) Temperature-dependent Debye temperature ($\Theta_\mathrm{D}$) of $\langle 100\rangle$ surface-oriented thin films of different thicknesses. Black circles denote a previous experiment \cite{Flubacher_1959}. (b) Thickness-dependent $\Theta_\mathrm{D}$ of thin films for three surface orientations. Plus markers represent $\Theta_\mathrm{D}$ of thin films calculated by the bulk-adapted method. Open triangles and circles denote $\Theta_\mathrm{D}$ for the $\langle 100\rangle$ and $\langle 111\rangle$ surface orientations, respectively, by excluding the contributions of the side peaks to the density of states (2.5--3.5 THz and 1.5--2.5 THz for the $\langle 100\rangle$ and $\langle 111\rangle$  surface orientations, respectively). The horizontal line indicates $\Theta_\mathrm{D}^\mathrm{bulk}$, while the dashed lines denote the fitting results of $\Theta_\mathrm{D}$ to the inverse of the thickness.}
\label{fig:5}
\end{figure}

Figure \ref{fig:5}(a) plots the temperature-dependent $\Theta_\mathrm{D}$ of the $\langle 100\rangle$ surface-oriented thin films for different thicknesses. Overall, $\Theta_\mathrm{D}$ increases monotonically as the temperature increases, and tends to converge at higher temperatures. In the following discussion, we use $\Theta_\mathrm{D}$ at $T$ = 1000 K. The thickness dependence of $\Theta_\mathrm{D}$ for the three surface orientations is presented in Fig. \ref{fig:5}(b). $\Theta_\mathrm{D}$ decreases monotonically with respect to the thickness; in particular, $\Theta_\mathrm{D}$ of the thinnest films is significantly lower than the bulk counterpart ($\Theta_\mathrm{D}^\mathrm{bulk}$ = 647 K) \cite{Flubacher_1959}. In contrast, $\Theta_\mathrm{D}$ of thin films calculated based on the bulk-adapted method is insensitive to the thickness regardless of the surface orientation, indicating that the presence of a surface is involved in determining the surface-oriented thickness dependence of $\Theta_\mathrm{D}$. Similar to the surface-to-volume ratio, $\Theta_\mathrm{D}$ for all surface orientations is inversely proportional to the thickness; however, the rate of convergence to $\Theta_\mathrm{D}^\mathrm{bulk}$ is dependent on the surface orientation. The difference in the convergence rate is primarily determined by the magnitude of harmonic IFC perturbations, more precisely the side peaks characterized by surface phonons. When excluding low-frequency side peaks of the $\langle 100\rangle$ and $\langle 111\rangle$ surface orientations, the calculated $\Theta_\mathrm{D}$ is not affected by the surface orientation and collapses onto the same thickness dependence. 

\begin{figure}[h]
\centering
\includegraphics[width = 0.8\columnwidth]{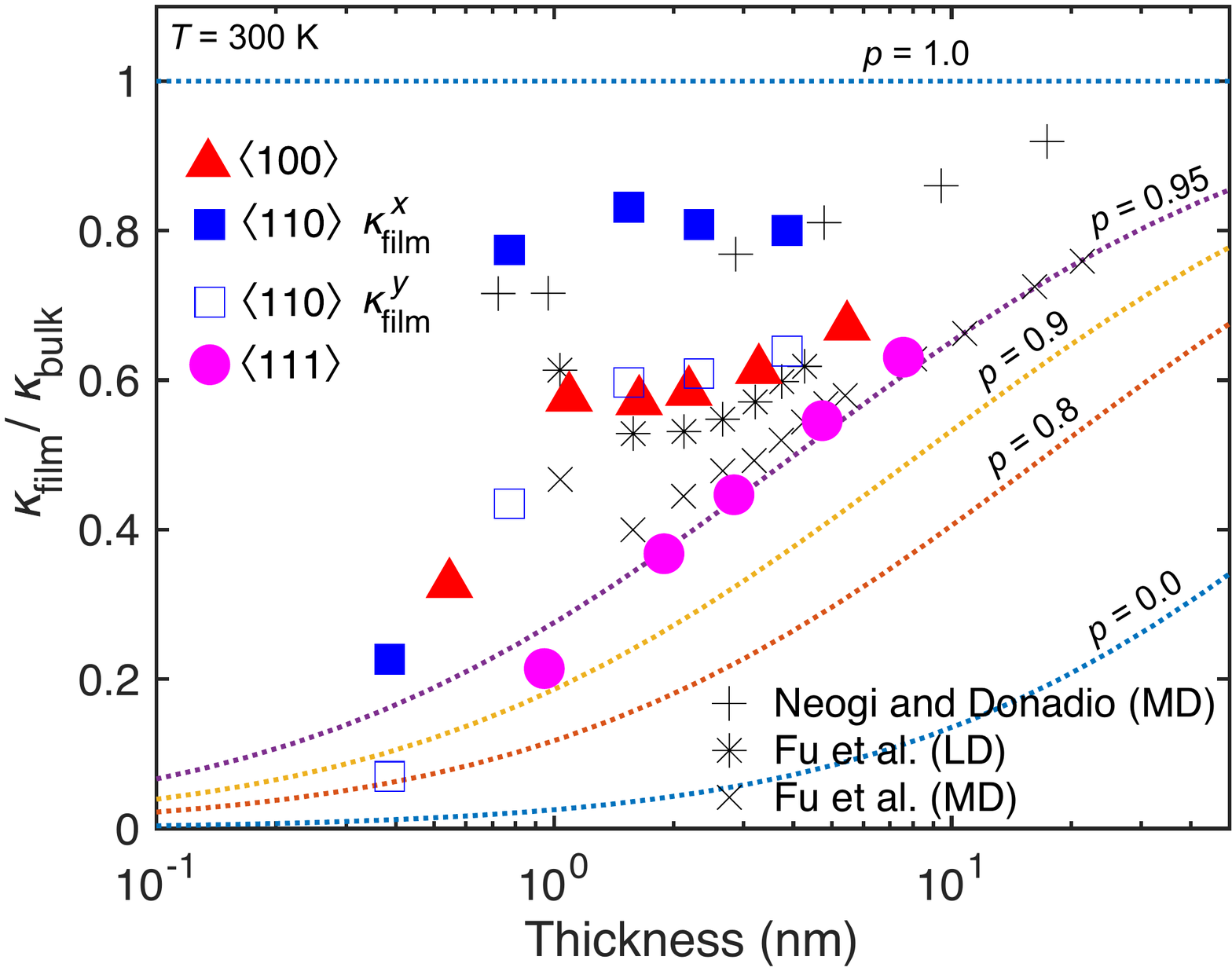} % reprint
\caption{Thickness dependence of in-plane thermal conductivity of thin films ($\kappa_\mathrm{film}$) normalized by the bulk thermal conductivity ($\kappa_\mathrm{bulk}$) at $T$ = 300 K. For the $\langle 110\rangle$ surface orientation, $\kappa_\mathrm{film}$ in the $x$- and $y$-directions ($\kappa_\mathrm{film}^x$ and $\kappa_\mathrm{film}^y$) is plotted due to the anisotropy of heat conduction. The dotted lines denote the Fuchs--Sondheimer  model \cite{Fuch_1938,Sondheimer_1952} with a given surface specularity ($p$). The markers denote the results of previous molecular dynamics (MD) and lattice dynamics (LD) calculations of $\kappa_\mathrm{film}$ \cite{Neogi_Donadio_2015,Fu_2020}.} 
\label{fig:6}
\end{figure}

Figure \ref{fig:6} displays the thickness dependence of the calculated in-plane thermal conductivity ($\kappa_\mathrm{film}$) at $T$ = 300 K for the three surface orientations. Due to the anisotropy of heat conduction, we plotted $\kappa_\mathrm{film}$ of the $\langle 110\rangle$ surface-oriented thin films in the $x$- and $y$-directions. Overall, $\kappa_\mathrm{film}$ decreases as the thickness decreases. For the $\langle 100\rangle$ surface orientation, the magnitude of the decrease in $\kappa_\mathrm{film}$ exhibits a plateau in the region of 1--2-nm thickness, which is consistent with previous calculations \cite{Neogi_Donadio_2015,Fu_2020,Turnery_2010}. When the thickness is below 1 nm, a further decrease in $\kappa_\mathrm{film}$ appears. This drastic reduction of $\kappa_\mathrm{film}$ can also be observed in other surface orientations, which can be attributed to the significant modulation of phonon dispersions for the thinnest films (Figs. \ref{fig:2} and \ref{fig:3}). Interestingly, for the $\langle 110\rangle$ surface orientation, $\kappa_\mathrm{film}$ in the $x$-direction ($\kappa_\mathrm{film}^x$) is nearly constant at thicknesses above 1 nm and is close to the bulk thermal conductivity ($\kappa_\mathrm{bulk}$). In contrast, $\kappa_\mathrm{film}$ in the $y$-direction ($\kappa_\mathrm{film}^y$) exhibits a similar thickness dependence as for the $\langle 100\rangle$ surface orientation. In the entire thickness regime, $\kappa_\mathrm{film}^y$ is lower than $\kappa_\mathrm{film}^x$ because several acoustic modes in the $\Gamma$--Y line are reduced and group velocities are lower than those in the $\Gamma$--X line (Fig. \ref{fig:3}(c)). 

For the $\langle 111\rangle$ surface orientation, $\kappa_\mathrm{film}$ is clearly different from that of the other surface orientations and is similar to the FS model with the surface specularity of $p$ = 0.95. As previous calculations have demonstrated the failure of the FS model, we also verified the validity of this model. In the FS model, surface specularity is often related to surface roughness \cite{Ziman,Maznev}. We estimated the surface roughness by fitting our results to the FS model, and compare it with the root-mean-square displacement of the outermost surface atoms, as our calculations did not include distinct roughness. The results indicate that the calculated root-mean-square displacement is lower than the estimated surface roughness by one order of magnitudes. Consequently, the FS model cannot explain the physics underlying the thickness dependence of $\kappa_\mathrm{film}$ for thicknesses below 10 nm. Since the dependence of $\kappa_\mathrm{film}$ on the surface orientation is similar to the dependence of the Debye temperature on the surface orientation, we can speculate that surface phonons are involved in reducing $\kappa_\mathrm{film}$ depending on the surface orientation. 

\begin{figure}[h]
\centering
\includegraphics[width = 0.5\columnwidth]{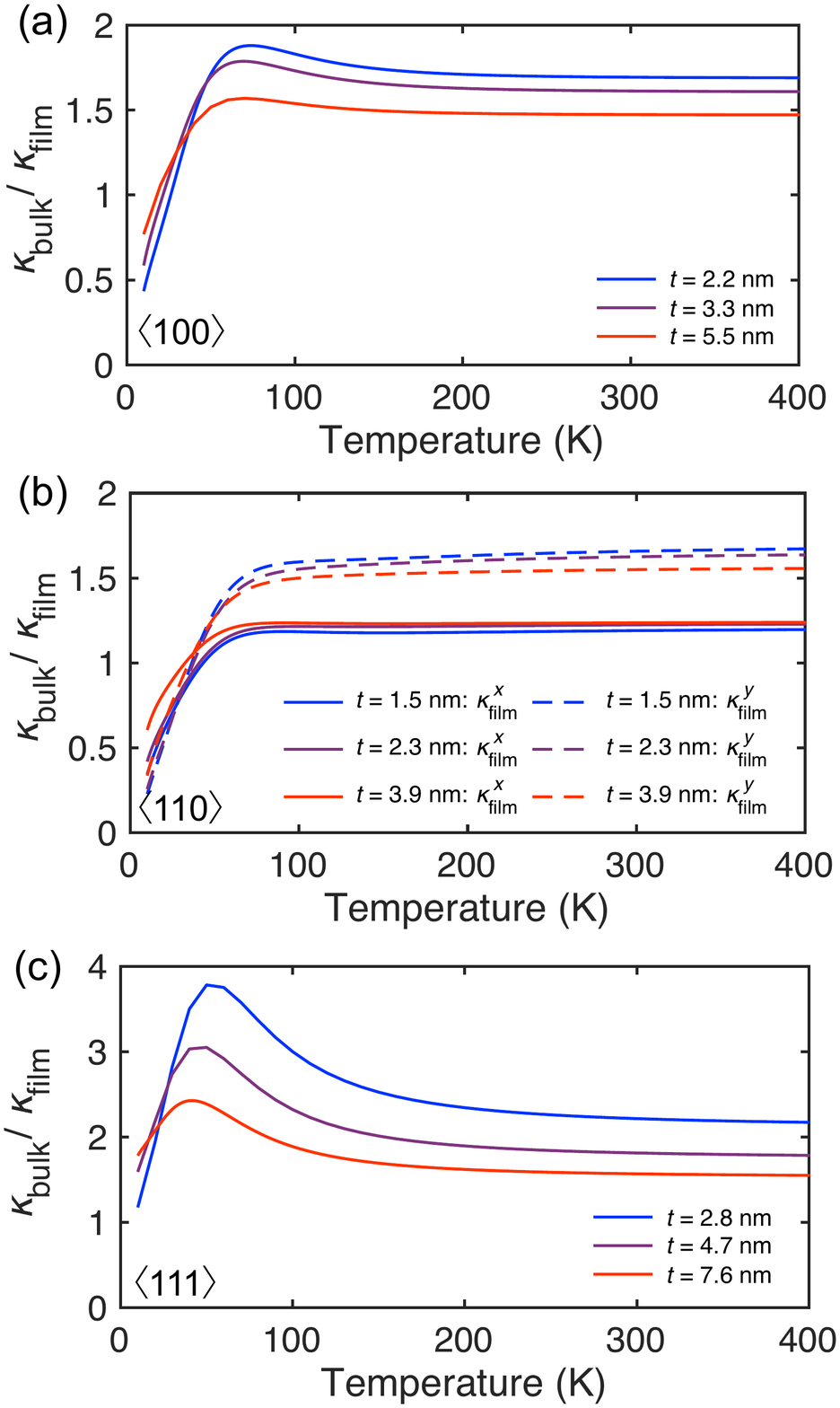} 
\caption{$\kappa_\mathrm{bulk}/\kappa_\mathrm{film}$ as a function of temperature for (a) $\langle 100\rangle$, (b) $\langle 110\rangle$, and (c) $\langle 111\rangle$ surface orientations. For the $\langle 110\rangle$ surface orientations, solid and dashed lines denote $x$- and $y$-directions, respectively.}
\label{fig:7}
\end{figure}

To gain more insight into how surface phonons qualitatively affect heat conduction in thin films, we calculated the temperature dependence of $\kappa_\mathrm{bulk}/\kappa_\mathrm{film}$ (Fig. \ref{fig:7}). Whereas $\kappa_\mathrm{bulk}/\kappa_\mathrm{film}$ for the $\langle 110\rangle$ surface orientation exhibits a monotonic trend, $\kappa_\mathrm{bulk}/\kappa_\mathrm{film}$ for the $\langle 100\rangle$ and $\langle 111\rangle$ surface orientations increases and then decreases as the temperature increases, and finally converges at higher temperatures. Although the temperature at which $\kappa_\mathrm{bulk}/\kappa_\mathrm{film}$ is the highest changes slightly depending on the thickness, it can be roughly estimated as 60--80 K and 40--60 K for the $\langle 100\rangle$ and $\langle 111\rangle$ surface orientations, respectively. When converting these temperature into the corresponding frequencies through $\omega~\sim~k_\mathrm{B}T/\hbar$, the converted frequencies are in reasonable agreement with those of the S$_1$ mode for the $\langle 100\rangle$ surface orientation and the S$_1$ and S$_2$ modes for the $\langle 111\rangle$ surface orientation (Figs. \ref{fig:2} and \ref{fig:3}). Therefore, the temperature dependence for the $\langle 100\rangle$ and $\langle 111\rangle$ surface orientations can be explained in terms of the thermal excitation of surface phonons as follows. Below the temperature at which surface phonons are thermally excited, the population of surface phonons increases as the temperature increases, suggesting that the influence of surface phonons on overall heat conduction in thin films is relatively large in the low-temperature regime. As the temperature further increases, since other phonons are thermally excited and participate in heat conduction, the proportion of surface phonons to all phonons becomes saturated, and the influence of surface phonons gradually decreases. 

Our results for the thickness and temperature dependence suggest that surface phonons have an impact on heat conduction in thin films. Here, we evaluate how three--phonon scattering involving surface phonons and other phonons localized in thin films (referred to as internal phonons) influences $\kappa_\mathrm{film}$. Three--phonon scattering processes fall into three groups: (i) scattering with only internal phonons, (ii) scattering with only surface phonons, and (iii) scattering involving internal and surface phonons. Of the three groups, we neglected the (iii) scattering processes (i.e., surface--internal phonon scattering) in the calculations of $\kappa_\mathrm{film}$. 

To verify our hypothesis, it is necessary to decompose surface and internal phonons. To this end, we applied the atomic participation ratio (APR) \cite{Hafner,Pailhes} to quantitatively decompose the surface and internal phonons. For a phonon mode with a wavevector $\bm{q}$ and polarization $s$, $F_{\bm{q}s}^\mathrm{APR}(i)$ indicates how the eigenvector of phonon mode $\bm{q}s$ is localized at the $i$th atom in a primitive unit cell, given by
\begin{equation}
	\displaystyle F_{\bm{q}s}^\mathrm{APR}(i) = \sqrt{N}\frac{|\bm{e}_{\bm{q}s}(i)|^{2}}{M_{i}}\left(\sum_{j}^{N}\frac{|\bm{e}_{\bm{q}s}(j)|^4}{M_{j}^{2}}\right)^{-1/2},
	\label{eq:APR}	
\end{equation}
where $i$ and $N$ denote the atomic index and number of atoms in a primitive unit cell, respectively. $\bm{e}_{\bm{q}s}(i)$ is the eigenvector of the $i$th atom of phonon mode $\bm{q}s$, and $M_{i}$ is the mass of the $i$th atom. $F_{\bm{q}s}^\mathrm{APR}(i)$ is unity when phonon mode $\bm{q}s$ is completely localized at the $i$th atom; otherwise, it is $1/\sqrt{N}$ for complete delocalization. Because there are two surfaces at the top and bottom of the thin film, we used $F_{\bm{q}s}^\mathrm{APR}(i)$ for the two outermost surface atoms in the decomposition. Here, the outermost surface is defined as the surface in contact with the vacuum layer. The hybridization of the surface and internal vibrations depends on the phonon mode; therefore, the value of $F_{\bm{q}s}^\mathrm{APR}(i)$ for decomposing the surface phonons cannot be uniquely determined. Additionally, the surface atomic density and specific surface area depend on the surface orientation and thickness, respectively. Thus, a single threshold value for $F_{\bm{q}s}^\mathrm{APR}(i)$ may not be appropriate. Nevertheless, we employed a single threshold value $F_\mathrm{thr}^\mathrm{APR}$ and set it to 0.3, which is reasonable for the decomposition of surface phonons (Appendix \ref{appendix:B}). It should be noted that we did not perform decomposition for thin films with thicknesses below 2.2 nm, 1.5 nm, and 2.8 nm for the $\langle 100\rangle$, $\langle 110\rangle$, and $\langle 111\rangle$ surface orientations, respectively, because surface and internal phonons are strongly hybridized and cannot be separated.

\begin{figure}[h]
\centering
\includegraphics[width = 0.8\columnwidth]{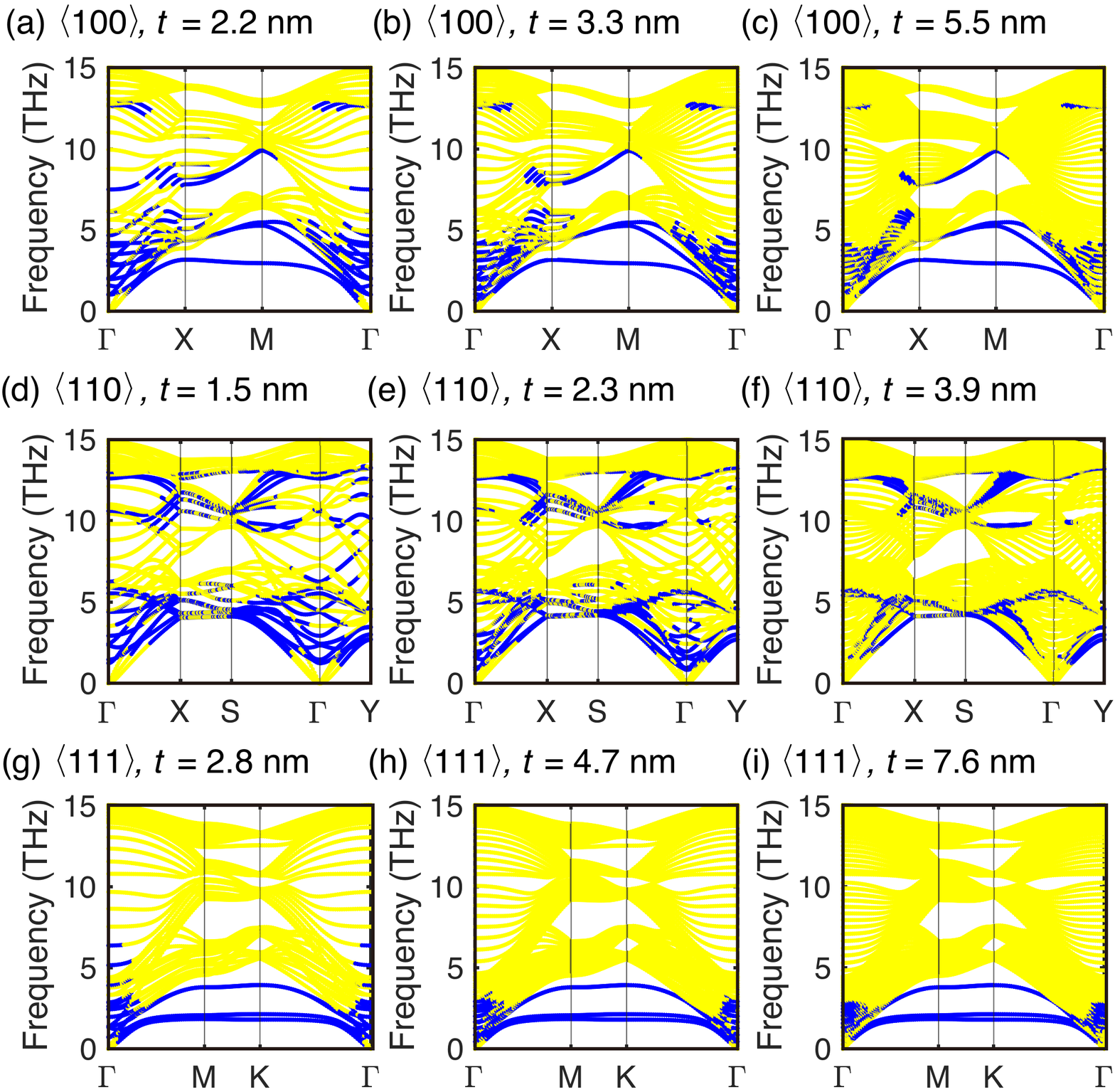} 
\caption{Atomic participation ratio projected onto phonon dispersion relations for (a)--(c) $\langle 100\rangle$, (d)--(f) $\langle 110\rangle$, and (g)--(i) $\langle 111\rangle$ surface-oriented thin films of different thicknesses. The blue and yellow colors denote surface and internal phonons, respectively.}
\label{fig:8}
\end{figure}

Figure \ref{fig:8} depicts color maps of $F_{\bm{q}s}^\mathrm{APR}(i)$ for the outermost surface atoms projected onto the dispersion relations of the $\langle 100\rangle$, $\langle 110\rangle$, and $\langle 111\rangle$ surface-oriented thin films. For the $\langle 100\rangle$ and $\langle 111\rangle$ surface orientations, although the single threshold value for $F_\mathrm{thr}^\mathrm{APR}$ failed to decompose part of the high-frequency surface phonon modes, several surface phonon modes characterized in each surface orientation can be successfully identified. Furthermore, some optical branches can also be identified as surface phonons. Interestingly, for the $\langle 110\rangle$ surface orientation, surface phonons, which are absent from the dispersion relations (Fig. \ref{fig:3}), can also be observed, and the frequencies of low-frequency surface phonons are consistent with those at which modulations in DOS spectra are observed (Fig. \ref{fig:4}). By examining the low-frequency regime, a common feature of three surface orientations in Fig. \ref{fig:8} is that surface phonons are identical to the acoustic modes except at the zone center. As the thickness increases or the surface-to-volume ratio decreases, the number of internal phonons naturally increases; however, some acoustic phonons close to the zone boundary are still classified as surface phonons.

\begin{figure}[h]
\centering
\includegraphics[width = 0.5\columnwidth]{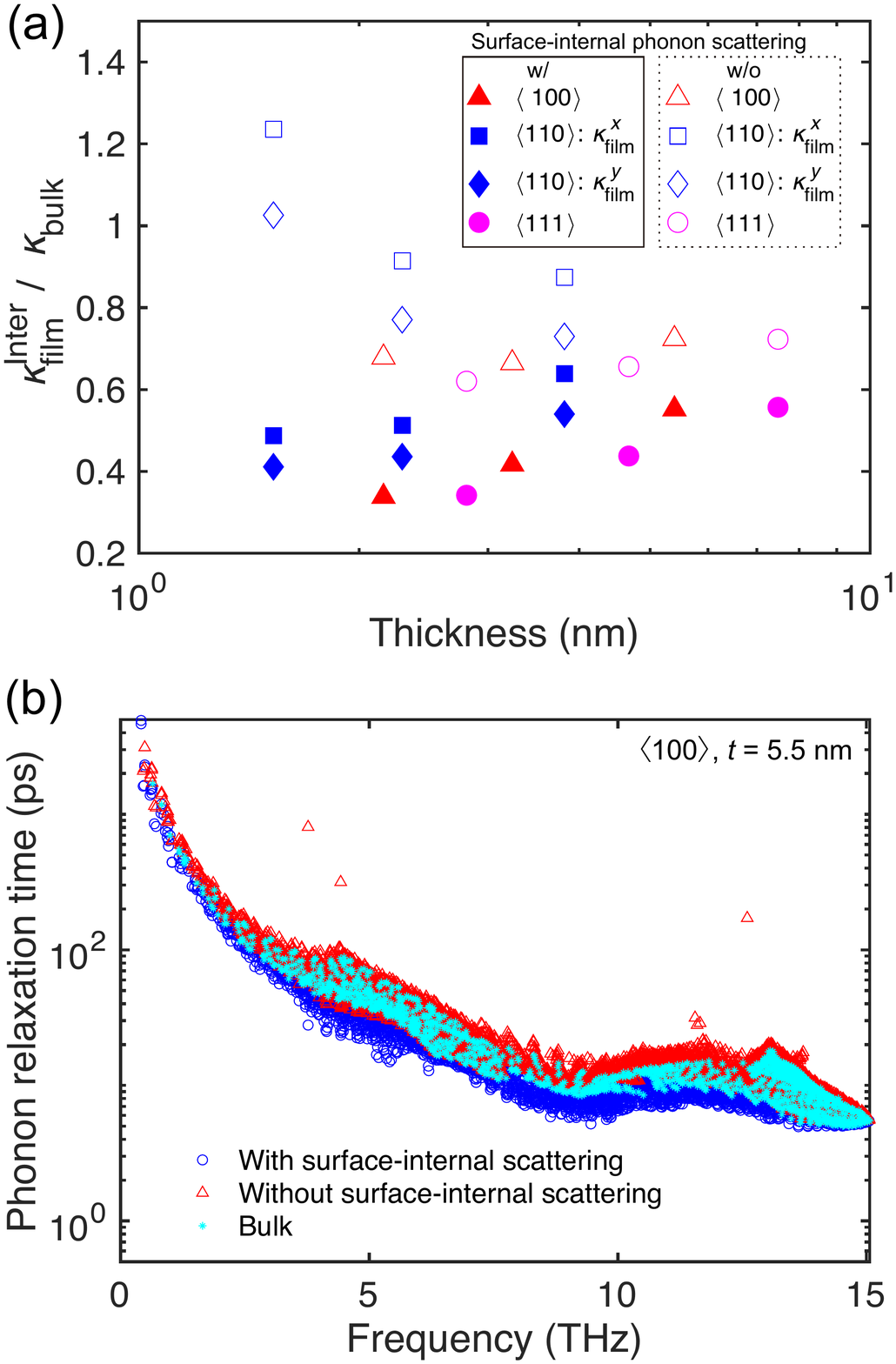} 
\caption{(a) Thickness-dependent in-plane thermal conductivity of internal phonons ($\kappa_\mathrm{film}^\mathrm{inter}$) at $T$ = 300 K for three surface orientations normalized by $\kappa_\mathrm{bulk}$ at the same temperature. Filled and opened markers represent $\kappa_\mathrm{film}^\mathrm{inter}$ with and without surface--internal phonon scattering, respectively. (b) Frequency-dependent relaxation times of internal phonons at $T$ = 300 K for the $\langle 100\rangle$ surface-oriented thin film of 5.5-nm thickness. Blue and red markers denote the relaxation times of internal phonons with and without surface--internal phonon scattering, respectively. Cyan markers represent the phonon relaxation times of bulk silicon at $T$ = 300 K for comparison.}
\label{fig:9}
\end{figure}

As the decomposition of surface and internal phonons was successful, we neglected surface-internal phonon scattering and calculated the thickness-dependent in-plane thermal conductivity of internal phonons ($\kappa_\mathrm{film}^\mathrm{inter}$) for three surface orientations at $T$ = 300 K, illustrated in Fig. \ref{fig:9}(a). The results for the $\langle 100\rangle$ surface orientation (red-opened triangles) indicate that the absence of surface-internal phonon scattering not only increases $\kappa_\mathrm{film}^\mathrm{inter}$, but also changes its thickness dependence compared to Fig. \ref{fig:6}. A similar result can be observed for the $\langle 111\rangle$ surface orientation. To gain further insight, we calculated the phonon relaxation times of the internal phonons for the $\langle 100\rangle$ surface-oriented 5.5-nm-thick film (Fig. \ref{fig:9}(b)). By neglecting surface-internal phonon scattering, the relaxation times of internal phonons are up to 1.5 times higher than that of their bulk counterparts. Another remarkable feature is the change in the frequency dependence of the relaxation times; namely, the absence of surface-internal phonon scattering makes the frequency dependence close to that of the bulk counterparts, indicating that surface--internal phonon scattering are dominant in three--phonon scattering and suppresses heat conduction of internal phonons. In addition, surface--internal phonon scattering also significantly hinders surface phonon transport (see Appendix \ref{appendix:C}). For the $\langle 110\rangle$ surface orientation, $\kappa_\mathrm{film}^\mathrm{inter}$ of 1.5-nm thickness exceeds $\kappa_\mathrm{bulk}$, which is due to the difficulty in the decomposition of surface phonons. 

\begin{figure}[h]
\centering
\includegraphics[width = 0.5\columnwidth]{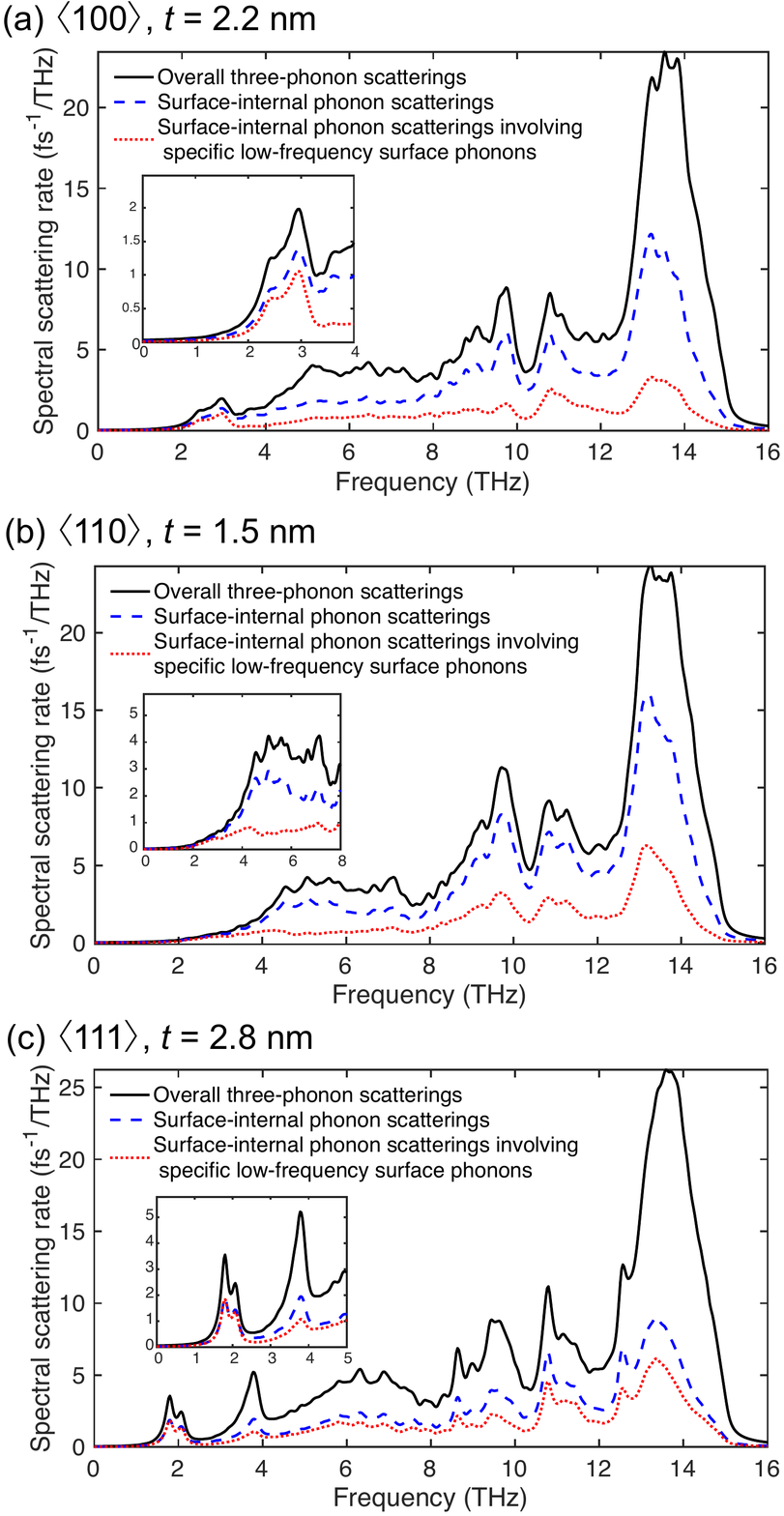} 
\caption{Spectral scattering rates of overall three--phonon scattering and surface--internal phonon scattering at $T$ = 300 K for (a) $\langle 100\rangle$ surface-oriented thin film of 2.2-nm thickness, (b) $\langle 110\rangle$ surface-oriented thin film of 1.5-nm thickness, and (c) $\langle 111\rangle$ surface-oriented thin film of 2.8-nm thickness. The solid black line denotes the overall three--phonon scattering, while the blue dashed line denotes surface--internal phonon scattering. The red dotted line denotes surface--internal phonon scattering involving specific low-frequency surface phonons (S$_1$ mode for the $\langle 100\rangle$ surface orientation and S$_1$ and S$_2$ modes for the $\langle 111\rangle$ surface orientation). The insets display enlarged regions of each graph.}
\label{fig:10}
\end{figure}

The temperature dependence (Fig. \ref{fig:7}) suggests that the low-frequency S$_1$ and S$_2$ surface phonon modes are involved in the reduced thermal conductivity for the $\langle 100\rangle$ and $\langle 111\rangle$ surface orientations; therefore, we discuss how these specific surface phonon modes contribute to the suppression of heat conduction. Figure \ref{fig:10} displays the frequency dependence of the spectral scattering rates for surface--internal phonon scattering at $T$ = 300 K. It should be mentioned that S$_1$ and S$_2$ are defined as surface phonon modes with the lowest and second-lowest frequencies, respectively. These labels are consistent with the discussions on phonon dispersion relations (Figs. \ref{fig:2} and \ref{fig:3}). As seen in Fig. \ref{fig:10}, surface--internal phonon scattering involving S$_1$ or both S$_1$ and S$_2$ surface phonon modes (represented as red-dotted lines) is predominant in the overall surface--internal phonon scattering (blue-dashed lines) in the low-frequency regime.

\begin{figure}[h]
\centering
\includegraphics[width = 0.8\columnwidth]{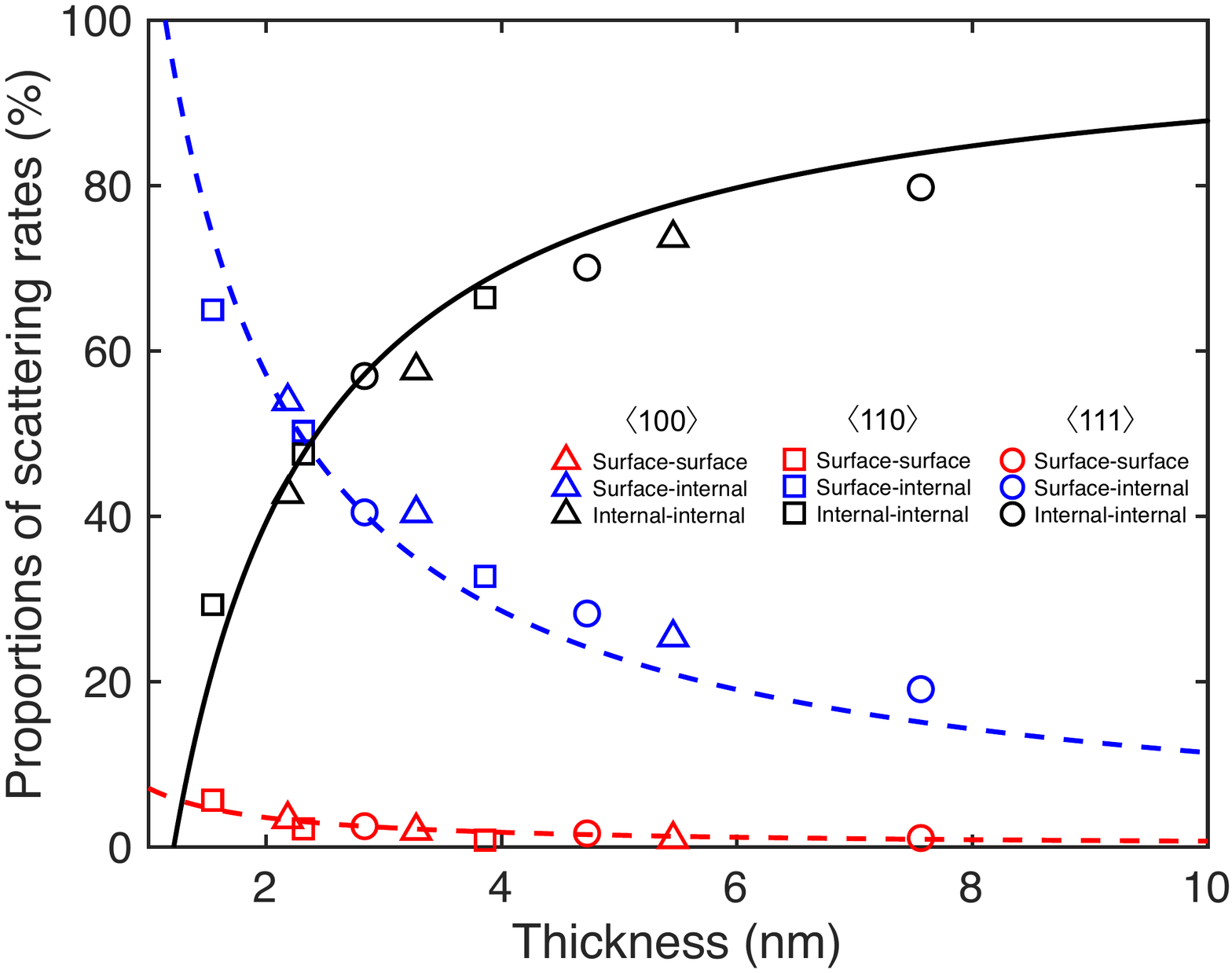} 
\caption{Thickness-dependent proportions of each scattering process to the overall three--phonon scattering at $T$ = 300 K for three surface orientations. Denoting the thickness as $t$, the dashed and solid lines denote the fitting results of the functions of $1/t$ and $100-1/t$, respectively.}
\label{fig:11}
\end{figure}

It is worth identifying which modes are coupled to the S$_1$ mode for the $\langle 100\rangle$ surface orientation and S$_1$ and S$_2$ modes for the $\langle 111\rangle$ surface orientation in the surface--internal phonon scattering. We thus investigated all triplets of the surface--internal phonon scattering and identified that the triplets of two S$_1$ surface phonons and one internal phonon in the vicinity of 6 THz (i.e., S$_1$ + S$_1~\rightarrow$ internal phonon and vice versa) account for 20\% of the overall surface--internal phonon scattering below 4 THz for the $\langle 100\rangle$ surface orientation. In contrast, for the $\langle 111\rangle$ surface orientation, triplets of the S$_1$ + S$_1~\rightarrow$ internal phonon, S$_2$~+~S$_2~\rightarrow$~internal phonon, and S$_1$~+~S$_2~\rightarrow$~internal phonon, and vice versa contribute to 40\% of the overall surface--internal phonon scattering below 3 THz. Unlike these two surface orientations, we did not find specific triplets for the $\langle 110\rangle$ surface orientation. As the surface-to-volume ratio decreases, the impact of surface--internal phonon scattering is expected to decrease monotonically. Figure \ref{fig:11} presents the proportions of surface--surface phonon scattering, surface--internal phonon scattering, and internal--internal phonon scattering to the overall three--phonon scattering for different thicknesses and surface orientations. The proportions of surface--surface phonon scattering and surface--internal phonon scattering are inversely proportional to the thickness, independent of surface orientation. By extrapolating the results, surface--surface phonon scattering and surface--internal phonon scattering account for 0.4\% and 5.7\% of the overall three--phonon scatterings, respectively, at a thickness of approximately 20 nm; thus, the impact of surface phonons should be limited in the sub-10-nm thickness regime and is negligible for heat conduction in thicker films. 

\section{Conclusion}
In the present work, we explicitly considered the atomic structures of thin films and performed anharmonic lattice dynamics calculations to investigate heat conduction in sub-10-nm-thick films. For harmonic properties, we observed that the presence of a surface not only leads to significant modulation of phonon dispersion relations, but also gives rise to surface phonons. The calculated thickness and temperature dependence of the in-plane thermal conductivity of thin films cannot be explained by conventional boundary scattering of phonons at surfaces, suggesting that the mechanism behind the significant suppression of heat conduction is affected by surface phonons. To investigate how surface phonon influence the suppression of heat conduction, we decomposed surface and internal phonons (localized in a thin film) from the perspective of the surface localization of vibrational modes. The results indicate that surface--internal phonon scattering predominantly influences the reduced thermal conductivity. Furthermore, we identified specific surface phonons and triplets in surface--internal phonon scattering that are dominant in the reduction of thermal conductivity. Since surface--internal phonon scattering can be enhanced or reduced by manipulating surface states through chemical functionalization and nanostructured surfaces, our findings can facilitate novel surface-phonon-engineered manipulation of heat conduction in thin films. 

\begin{acknowledgments}
This work was partially supported by PRESTO “Thermal Science and Control for Spectral Energy Transport” (Grant No. JPMJPR17I5) from the Japan Science and Technology Agency and a Grant-in-Aid for Scientific Research B (Grant No. 20H02080).  
\end{acknowledgments}

\appendix
\renewcommand\thefigure{\thesection\arabic{figure}} 
\setcounter{figure}{0}    
\section{\label{appendix:A}Details of calculations of in-plane thermal conductivity}
\begin{figure}[h]
\centering
\includegraphics[width = 0.5\columnwidth]{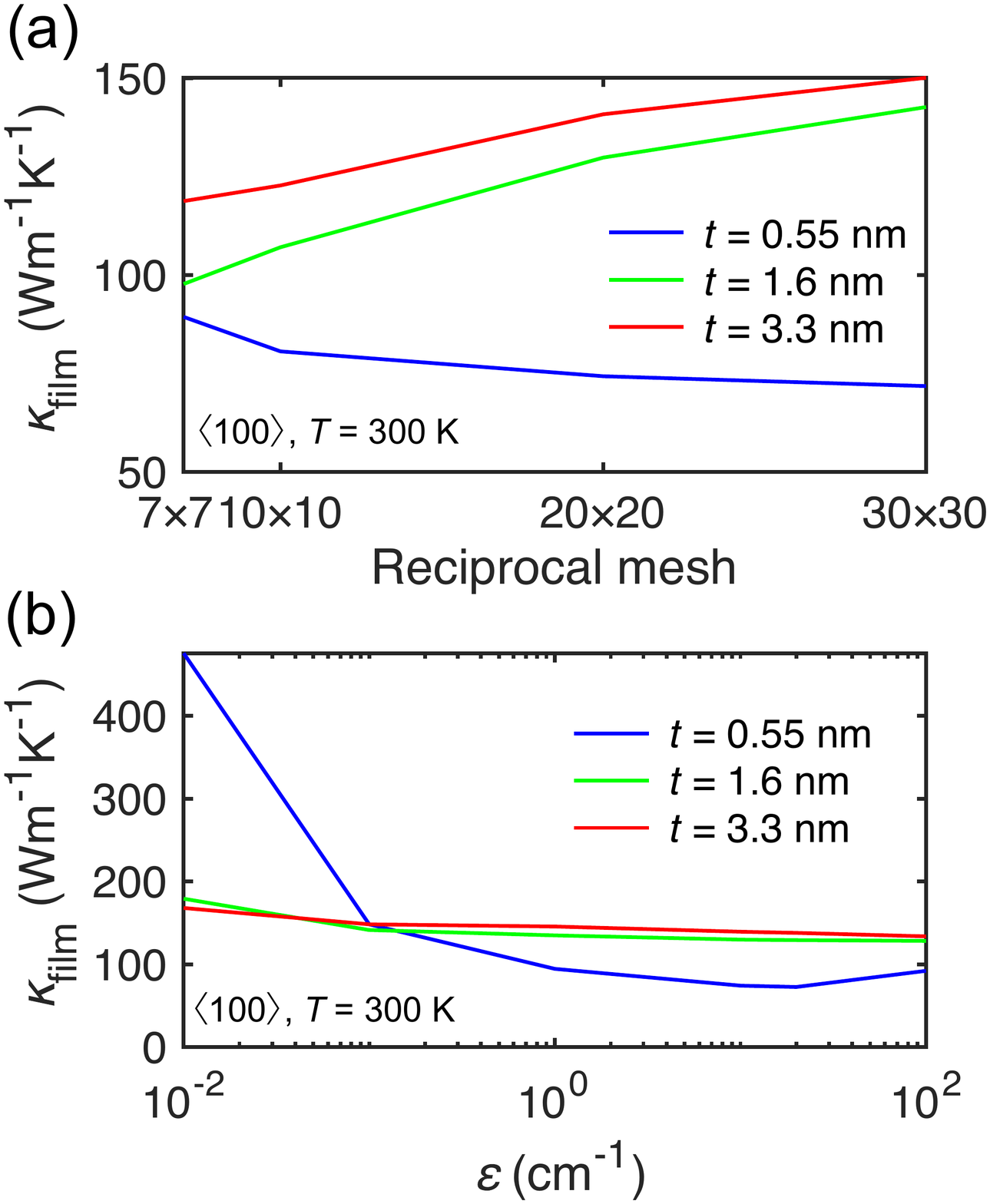} 
\caption{(a) Calculated in-plane thermal conductivity as a function of the uniform reciprocal mesh for $\langle 100\rangle$ surface-oriented thin films of three different thicknesses at $T$ = 300 K. (b) Dependence of in-plane thermal conductivity of $\langle 100\rangle$ surface-oriented thin films on the linewidth for three thicknesses. In the calculations, a $20~\times~20$ uniform reciprocal mesh was used.}
\label{fig:A1}
\end{figure}

Figure \ref{fig:A1}(a) plots the reciprocal mesh dependence of the in-plane thermal conductivities of $\langle 100\rangle$ surface-oriented thin films of three different thicknesses at $T$ = 300 K. Although the difference in thermal conductivity calculated with $20~\times~20$ and $30~\times~30$ uniform reciprocal meshes is at most approximately 10\%, we employed a $20~\times~20$ reciprocal mesh for all calculations, which is sufficient for discussing how surface phonons influence heat conduction in thin films. For the linewidth used in the calculations for three--phonon scattering, $\varepsilon$ = 10 cm$^{-1}$ chosen in our calculations is reasonable for the convergence of thermal conductivity (Fig. \ref{fig:A1}(b)). 

\setcounter{figure}{0} 
\section{\label{appendix:B}Sensitivity of the decomposition of surface phonons to the atomic participation ratio threshold value}
\begin{figure}[h]
\centering
\includegraphics[width = 0.8\columnwidth]{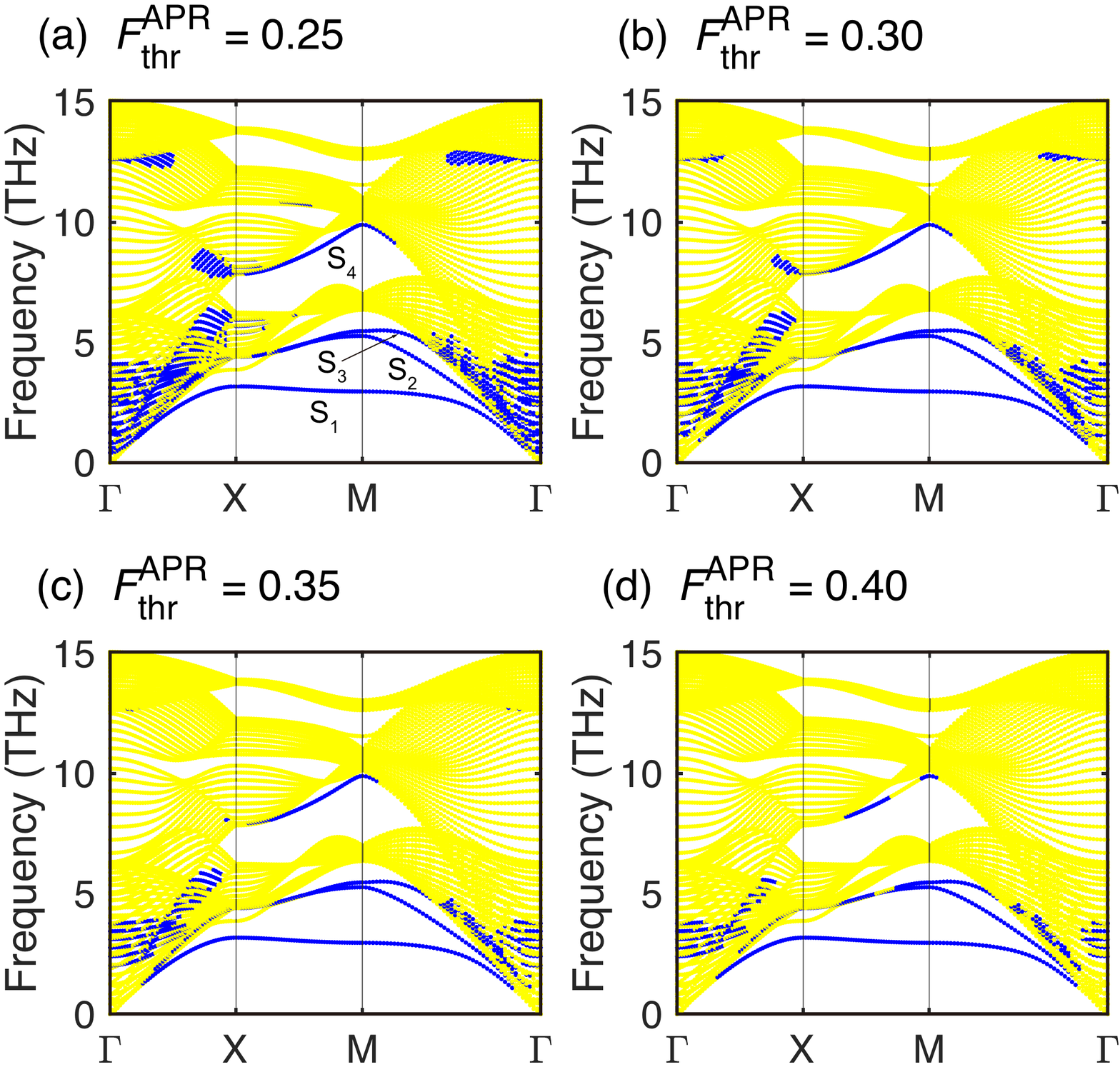} 
\caption{Atomic participation ratio projected onto the phonon dispersion relations of $\langle 100\rangle$ surface-oriented thin films of 5.5-nm thickness for different threshold values ($F_\mathrm{thr}^\mathrm{APR}$). Blue and yellow colors denote surface and internal phonons, respectively.}
\label{fig:B1}
\end{figure}

\begin{figure}[h]
\centering
\includegraphics[width = 0.8\columnwidth]{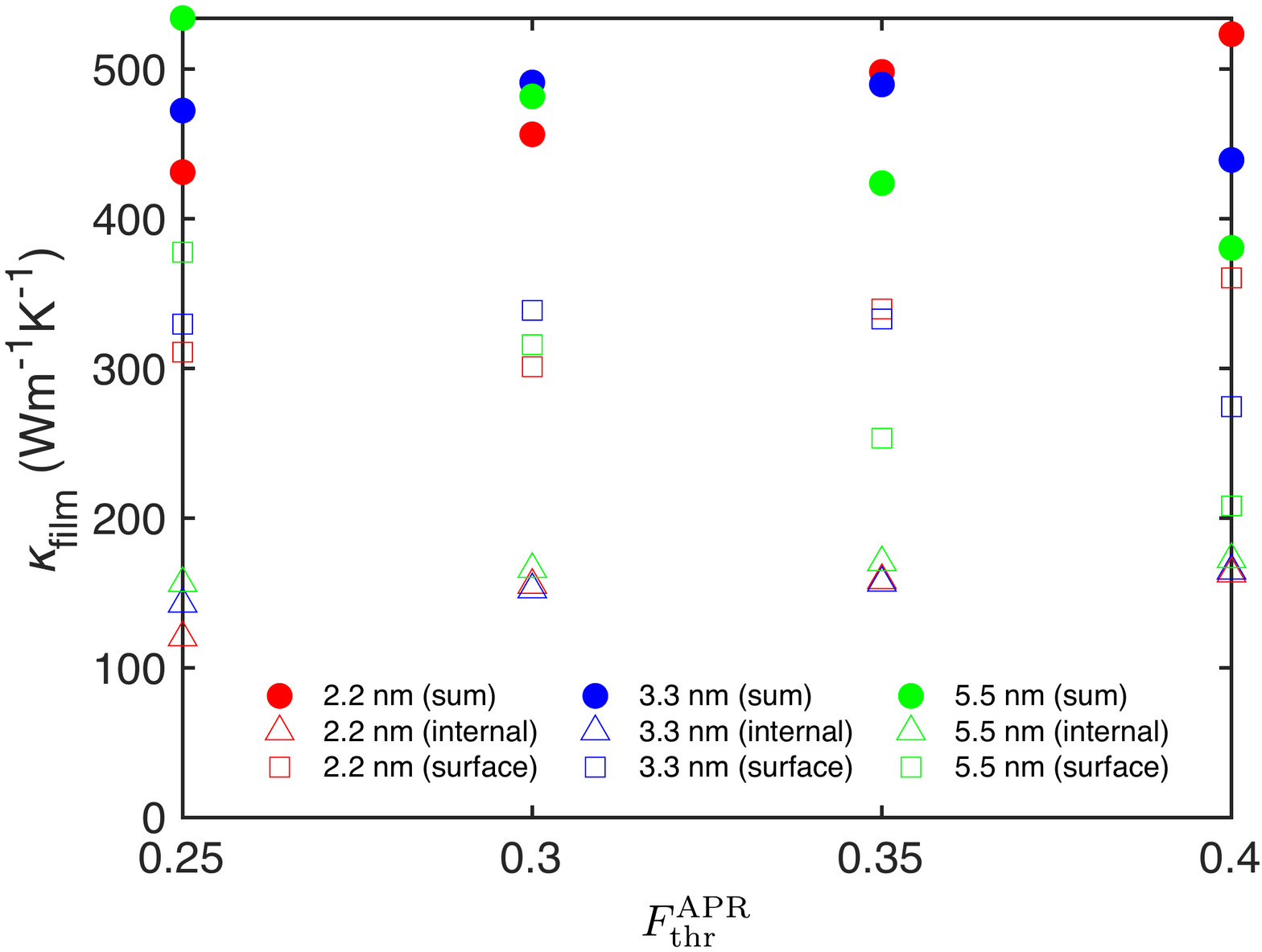} 
\caption{$F_\mathrm{thr}^\mathrm{APR}$-dependent in-plane thermal conductivity of surface and internal phonons ($\kappa_\mathrm{film}^\mathrm{surface}$, $\kappa_\mathrm{film}^\mathrm{inter}$, and the sum) of $\langle 100\rangle$ surface-oriented thin films at $T$ = 300 K for different thicknesses. Surface-internal phonon scattering is neglected in the calculations.}
\label{fig:B2}
\end{figure}

Figure \ref{fig:B1} illustrates the atomic participation ratio projected onto the phonon dispersion relations for the $\langle 100\rangle$ surface-oriented thin film of 5.5-nm thickness for different threshold values $F_\mathrm{thr}^\mathrm{APR}$. The number of decomposed surface phonons increases as $F_\mathrm{thr}^\mathrm{APR}$ decreases. Due to the strong localization of low-frequency surface phonon modes, the S$_1$--S$_3$ modes are robust to $F_\mathrm{thr}^\mathrm{APR}$. To investigate how $F_\mathrm{thr}^\mathrm{APR}$ influences heat conduction in thin films, we neglected surface--internal phonon scattering and calculated the in-plane thermal conductivity of surface and internal phonons ($\kappa_\mathrm{film}^\mathrm{surface}$ and $\kappa_\mathrm{film}^\mathrm{inter}$, respectively) at $T$ = 300 K (Fig. \ref{fig:B2}). The decomposition of surface and internal phonons significantly changes the magnitude of surface-internal phonon scattering and consequently results in the fluctuation of $\kappa_\mathrm{film}^\mathrm{surface}$. In contrast, $\kappa_\mathrm{film}^\mathrm{inter}$ is nearly independent of $F_\mathrm{thr}^\mathrm{APR}$. This feature can also be observed for other surface orientations. 

\setcounter{figure}{0} 
\section{\label{appendix:C}In-plane thermal thermal conductivity of surface phonons in the absence of surface--internal phonon scattering}
\begin{figure}[h]
\centering
\includegraphics[width = 0.5\columnwidth]{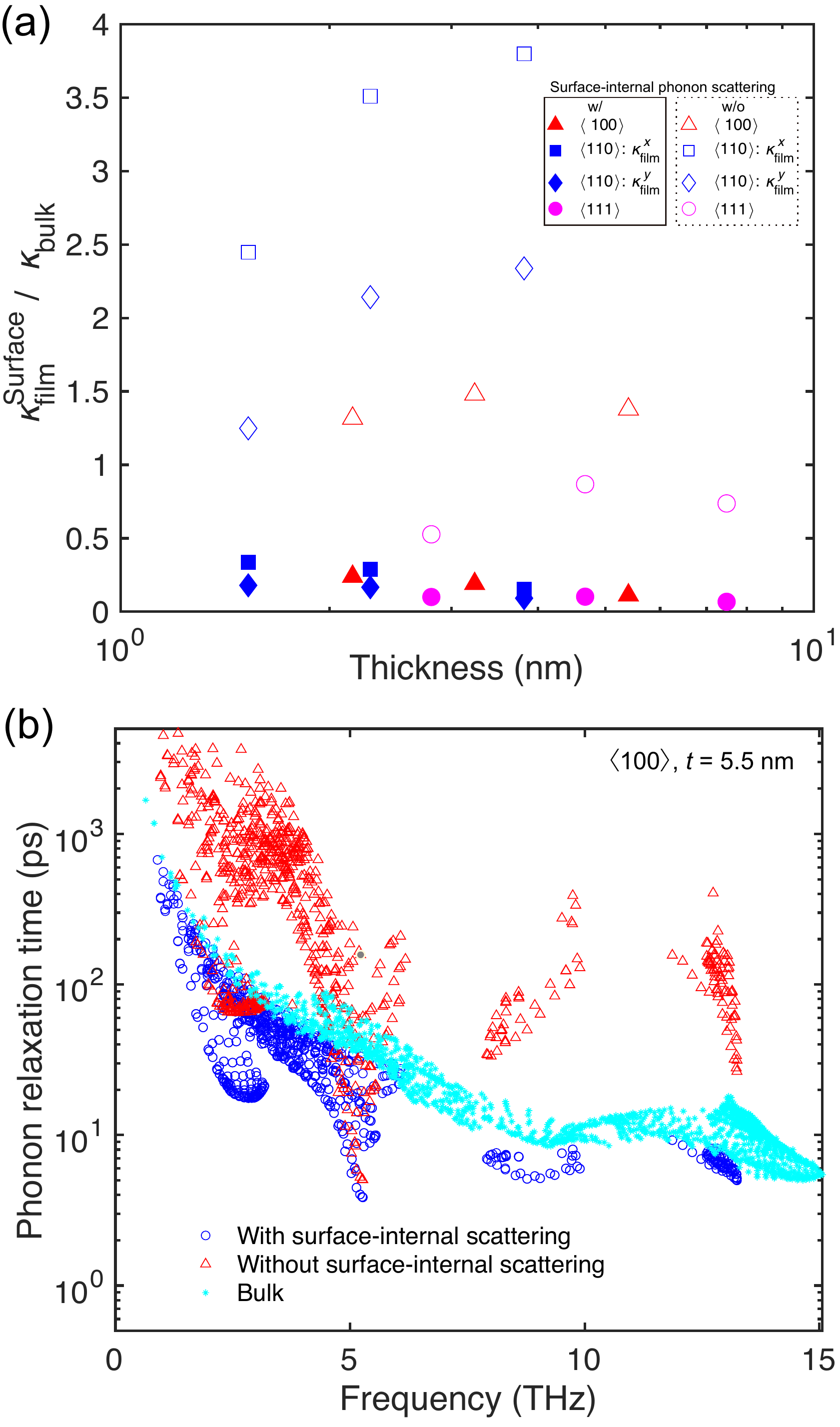} 
\caption{(a) Thickness-dependent $\kappa_\mathrm{film}^\mathrm{surface}$ with and without surface--internal phonon scattering at $T$ = 300 K for three surface orientations normalized by $\kappa_\mathrm{bulk}$ at the same temperature. (b) Frequency-dependent relaxation times of surface phonons at $T$ = 300 K for $\langle 100\rangle$ surface-oriented 5.5-nm-thick film. Blue and red markers denote the relaxation times with and without surface--internal phonon scattering, respectively. Cyan markers represent the relaxation times of bulk phonons for comparison.}
\label{fig:C1}
\end{figure}

Figure \ref{fig:C1}(a) displays the thickness dependence of $\kappa_\mathrm{film}^\mathrm{surface}$ with and without surface--internal phonon scattering in the case of $F_\mathrm{thr}^\mathrm{APR}$ = 0.3. For all surface orientations, similar to the results for internal phonons, the absence of surface--internal phonon scattering increases $\kappa_\mathrm{film}^\mathrm{surface}$. Due to the large surface-to-volume ratio, the proportion of surface phonons to all phonons is relatively large (approximately 30\%), which is one of the reasons for the high $\kappa_\mathrm{film}^\mathrm{surface}$. Another reason is the large relaxation times of surface phonons. Surface phonons are mainly coupled to internal phonons; thus, the scattering phase space \cite{Lindsay} for surface--surface phonon scattering is relatively small, resulting in a large increase in the relaxation times of surface phonons (Fig. \ref{fig:C1}(b)). 

\bibliography{reference}% Produces the bibliography via BibTeX.
\end{document}